\documentclass[preprintnumbers]{revtex4}

\usepackage{graphicx}
\def\ie{{\it i.e.}}
\def\eg{{\it e.g.}}

\def\etal{{\it et al.}}

\def\mpl{\ifmmode \overline M_{Pl}\else $\overline M_{Pl}$\fi}

\begin{document}
\bibliographystyle{revtex}

\preprint{SLAC-PUB-8973/
          P3-04}

\title{Distinguishing Kaluza-Klein Resonances From a $Z'$ in Drell-Yan 
Processes at the LHC}

\author{Thomas G. Rizzo}

\email[]{rizzo@slac.stanford.edu}
\affiliation{Stanford Linear Accelerator Center, 
Stanford University, Stanford, California 94309 USA}

\date{\today}

\begin{abstract}
We explore the capability of the LHC to distinguish the production of 
Kaluza-Klein(KK) excitations in Drell-Yan collisions from an ordinary $Z'$ at 
the LHC in the case of one extra dimension with the fermions localized at the 
orbifold fixed points. In particular, we demonstrate that this capability is 
dependent on both the mass of the KK state as well as whether or not the 
quarks and leptons lie at the same fixed points.
\end{abstract}

\maketitle

\section{Introduction: The Problem}

The possibility of KK excitations of the Standard Model(SM) gauge bosons 
within the framework of theories with TeV-scale extra dimensions has been 
popular for some time{\cite {anton}}. Given the many fields of the SM it is 
possible to construct a large number of interesting yet different models of 
this class depending on, \eg, whether all the gauge fields experience the same 
number of dimensions or whether the fermions and/or Higgs bosons 
are in the bulk. 
Perhaps the simplest model of this kind is the case of only one extra 
dimension where all of the SM gauge fields are in the bulk and the fermions 
lie at one of the two orbifold fixed points associated with the 
compactification on $S_1/Z_2${\cite {bunch}}. In this scheme the couplings of 
the KK excitations of a given gauge field are identical to those of the SM 
apart from an overall factor of 
$\sqrt 2$ and their masses are given, to lowest order in 
$(M_0/M_c)^2$, by the relationship $M_n^2=(nM_c)^2+M_0^2$, where $n$ labels the 
KK level, $M_c \sim 1$ TeV is the compactification scale and $M_0$ is the 
zero-mode mass obtained via spontaneous symmetry breaking for the cases of the 
$W$ and $Z$. Note for the 
cases of the photon and $Z$ that their first excitations will be highly 
degenerate in mass, becoming more so as $M_c$ increases. For example, if 
$M_c=4$ TeV the splitting between the $Z$ and $\gamma$ KK states is less than 
$\sim$2.5 GeV. An updated analysis{\cite {bunch}}  
of precision electroweak data implies that $M_c \geq 4$ TeV, independently of 
the location of the Higgs field, which is in a range accessible to the LHC. Of 
course this implies that 
the LHC experiments will at best observe only a {\it single} bump in the 
$\ell^+\ell^-$ channel and a corresponding single Jacobian peak in $\ell^\pm+$ 
missing $E$ channel as the next set of KK states 
is too massive to be seen even with 
an integrated luminosity of $100-300$ $fb^{-1}${\cite {tgr}}. 

How will this observation be interpreted? Through straightforward measurement 
of the lepton pair angular distribution it will be known immediately that the 
resonance is spin-1 and not, \eg, a spin-2 graviton resonance as in the 
Randall-Sundrum{\cite {rs}} 
model{\cite {dhr}}. Perhaps the most straightforward 
possibility is that of an extended gauge model{\cite {snow}} which predicts the 
existence of a degenerate $W'$ and $Z'$; many such models exist in the 
literature{\cite {models}}. Is it possible to distinguish this $Z'/W'$ 
model from KK excitations? In earlier work{\cite {tgr}} it was 
demonstrated that 
once the mass of the first KK excitation was determined at the LHC, a 
linear collider(LC)  
with an integrated luminosity of order 300 $fb^{-1}$ and a center of mass 
energy of 0.5(1) TeV could be used to distinguish the two scenarios for KK 
masses as high as $\simeq$ 5(7) TeV by examining how such new states would 
modify fermion pair production cross sections and asymmetries. (Note that these 
measurements are taking place far below the actual mass of the new 
excitation.)  The question 
we would like to address here is whether or not one has to wait for the LC in 
order to make this distinction, \ie, what can be done at the LHC itself? Can 
measurements at the LHC distinguish the two scenarios? We report here the 
preliminary results of a first analysis designed to address this issue.

\begin{figure}[htbp]
\centerline{
\includegraphics[width=7cm,angle=90]{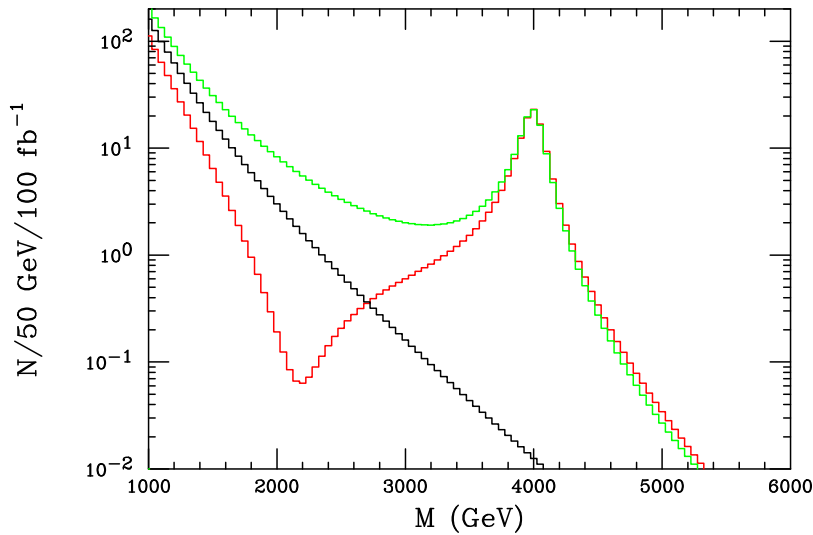}}
\vspace*{0.1cm}
\caption{Binned $\mu^+\mu^-$ Drell-Yan mass spectrum for the SM (black) and 
for the case 
of a 4 TeV KK excitation when all fermions are at the same orbifold fixed 
point(red) and when quarks and leptons are at opposite fixed points(green) thus 
separated by a distance $D=\pi R_c$ in the extra dimension. Rapidity cuts, 
K-factors and efficiencies are included.}
\label{p3-04_fig1}
\end{figure}

\section{Analysis: The Solution}

The only possible approach to this problem is to make precision measurements 
of the lepton pair 
invariant mass distribution. To get an idea of what this distribution 
would look like we show a representative example in Fig.~\ref{p3-04_fig1} 
for the case when $M_c=4$ TeV. (This very closely resembles the same plot 
after being put through a fast ATLAS detector simulation{\cite {giac}} giving 
us some confidence in our numerical study below.) 
Here we consider two cases: ($i$) all SM fermions 
are at the same orbifold point($D=0$) and ($ii$) quarks and leptons are 
at opposite 
fixed points separated by a distance $D=\pi R_c$ in the extra dimension where 
$R_c=1/M_c$. The later model may be of interest in addressing, \eg, the issue 
of proton decay. Note that in the LC analysis the value of $D$ did not enter  
since only leptonic data was employed. Here one may easily 
imagine that the capability 
of the LHC to distinguish the $Z'$ and KK scenarios may depend on $D$. Note 
that for $D=0(\pi R_c)$ there is a strong destructive(constructive) 
interference between the SM and KK contributions. 

What portion of the lepton mass 
spectrum is useful for this analysis? The resonance peak 
region is {\it not} useful (at least by itself) since, as many earlier 
$Z'$ analyses have shown{\cite {snow}}, for such a heavy $Z'$-like state 
the only useful data obtainable there 
are the total cross section, the full width and the 
forward-backward asymmetry, $A_{FB}$. 
The first two of these 
are sensitive to other potential non-SM decay modes and are thus 
highly model dependent while $A_{FB}$ is insufficient as a useful 
discriminator. 
Beyond the peak region the cross section is quite small yielding too poor a set 
of statistics to be valuable; this implies 
that only the low mass range is useful. To be specific we first generate 
Drell-Yan $\mu^+\mu^-$-pair cross section `data' for both the $D=0$ and 
$D=\pi R_c$ cases integrated over 100 GeV wide mass bins 
covering a dilepton mass region between 250 GeV and 1850(2150) GeV for the case 
of $M_c=4(5)$ TeV with an assumed integrated luminosity of 300 $fb^{-1}$. (To 
go lower in mass would not be very useful as we are then dominated by either 
the $Z$ peak or the photon pole. For larger masses the cross section is either 
too small or is dominated by the heavy resonance.) Next, under the assumption 
that a $Z'$ of known mass is actually being produced, we vary it's couplings 
in order to obtain the best $\chi^2$ fit to the dilepton mass 
distribution and obtain the 
relevant confidence level(CL) for the fit. (In this approach, the overall 
normalization of the cross section is determined at the $Z$-pole which is 
outside of the fit region.) In performing this analysis we make 
the following simplifying assumptions: ($i$) the $Z'$ couplings are generation 
independent and ($ii$) the generator to which the $Z'$ couples commutes with 
those of the SM. These conditions are satisfied by GUT-inspired $Z'$ models 
as well as by 
many others in the literature{\cite {snow}} and reduces the number 
of fit parameters to 5: the left-handed couplings of the quark and lepton 
doublets and the right-handed couplings for $u,d$ and $e$. We then perform a 
fine-grained scan over a large volume of this parameter space testing more 
than $10^{10}$ coupling combinations for each of the cases we consider to 
obtain the best fit. 

The results of this analysis are as follows. For the most naive case, where 
all the SM fermions are at the same orbifold fixed point, \ie, $D=0$, we find 
that the largest value of the CL obtained by our fitting procedure to be 
$\sim 10^{-10}(0.003)$ for the case $M_c=4(5)$ TeV. This implies that the 
assumption that the KK state is actually a $Z'$ does not provide a good fit 
and we can conclude that the two cases are distinguishable. However as we 
clearly see the CL of the fit in this case rises rapidly as $M_c$ 
increases since the influence of the resonance in the below peak region to 
which we are fitting 
is rapidly diminishing. For $M_c=6$ TeV CL's in the  range 
$0.5-1$ are easily obtained and the two scenarios are no longer separable. The 
results for the case $D=\pi R_c$ are quite different from those for $D=0$ 
since there is now constructive interference between the SM and KK 
contributions. In this case for $M_c=4(5)$ TeV the CL of the fits ranged as 
high as $\simeq$ 0.7(1) implying very good fits to the KK data with the $Z'$ 
hypothesis were possible even for relatively light masses. This implies that 
in this case the LHC will not be able to distinguish the KK and $Z'$ cases when 
the quarks and leptons are not at the same fixed points. This is seen to hold 
true for any value of $M_c$ which is  
in excess of the current bounds from precision electroweak data.

\section{Summary and Conclusions}

The identification of new physics after its discovery is an important issue 
for both present and future colliders. In the preliminary 
analysis presented above we 
considered the capability for the LHC to distinguish a KK excitation from a 
more conventional $Z'$ in the mass range at and 
above 4 TeV. Earlier analyses have 
shown that such a model separation is possible at a LC running at a fixed 
center of mass energy provided the mass of 
the excitation is already known from LHC measurements. In the case of the LHC 
we demonstrated that in the most naive scenario where all of the SM fermions 
are located a single orbifold fixed point the LHC is able to distinguish the 
two scenarios up to KK excitation masses in the 5-6 TeV range. On the 
otherhand, in the case where the quarks and leptons are at different fixed 
points, we have found that the LHC would find the two scenarios to be 
indistinguishable. It is possible that some extension of the current analysis 
may lead to a strengthening of the LHC's ability at model discrimination; 
this is currently under investigation. A detector simulation along the lines 
of the present analysis would be highly useful in verifying our results.

%
%%%%%%%%%%%%%%%%%%--- References
%%%%%%%%%%%%%%%%%%%%%%%%%%%%%%%%%%%%%%%%%%%%%%%%%%%%%%%
\def\MPL #1 #2 #3 {Mod. Phys. Lett. {\bf#1},\ #2 (#3)}
\def\NPB #1 #2 #3 {Nucl. Phys. {\bf#1},\ #2 (#3)}
\def\PLB #1 #2 #3 {Phys. Lett. {\bf#1},\ #2 (#3)}
\def\PR #1 #2 #3 {Phys. Rep. {\bf#1},\ #2 (#3)}
\def\PRD #1 #2 #3 {Phys. Rev. {\bf#1},\ #2 (#3)}
\def\PRL #1 #2 #3 {Phys. Rev. Lett. {\bf#1},\ #2 (#3)}
\def\RMP #1 #2 #3 {Rev. Mod. Phys. {\bf#1},\ #2 (#3)}
\def\NIM #1 #2 #3 {Nuc. Inst. Meth. {\bf#1},\ #2 (#3)}
\def\ZPC #1 #2 #3 {Z. Phys. {\bf#1},\ #2 (#3)}
\def\EJPC #1 #2 #3 {E. Phys. J. {\bf#1},\ #2 (#3)}
\def\IJMP #1 #2 #3 {Int. J. Mod. Phys. {\bf#1},\ #2 (#3)}
\def\JHEP #1 #2 #3 {J. High En. Phys. {\bf#1},\ #2 (#3)}

\end{document}